# Attosecond quantum optics

Mohamed Sennary[1,2†], Javier Rivera-Dean[3,4†], Yihe Wang[1,2], Maciej Lewenstein[3], Mohammed Th. Hassan[1,2*].

[1]Department of Physics, University of Arizona, Tucson, AZ 85721, USA.

[2] James C. Wyant College of Optical Sciences, University of Arizona, Tucson, Arizona 85721, nm'USA.

[3] ICFO–Institut de Ciencies Fotoniques, The Barcelona Institute of Science and Technology, Castelldefels (Barcelona) 08860, Spain.

[4]Department of Physics & Astronomy, University College London Gower Street London WC1E BT, United Kingdom.

*Corresponding author. Email: mohammedhassan@arizona.edu.

*†These authors contributed equally to this work.*

**Abstract:**

Modern quantum optics primarily operates in the quasi-stationary regime, isolated from the intrinsic timescales of ultrafast optical fields. Pushing these boundaries into the femtosecond and attosecond domains is a critical frontier. Here, we generate, shape, and interrogate the quantum state of an ultrafast squeezed light field. Our optical metrology reveals a highly dynamic, time-dependent squeezing distribution across individual half-cycles of the electric field. Incorporating this intra-cycle squeezing into strong-field simulations demonstrates that the temporal redistribution of quantum uncertainty fundamentally reshapes the quantum strong field physics of high-harmonic emission. Furthermore, we achieve attosecond-scale control of the squeezed state, visualized through inferred effective Wigner representations. Finally, we show that ultrafast squeezed light encodes its quantum properties into a photoinduced tunneling current within a petahertz phototransistor with sub-femtosecond resolution, demonstrating a direct optical–electronic quantum coupling. This work lays the foundation for the emerging field of ultrafast quantum optics and unlocks new avenues for high-speed quantum communication and photonics.

Quantum optics fundamentally relies on controlling the quantum state of light. Squeezed states of light, which is a nonclassical light exhibiting reduced noise in one quadrature at the expense of another, have driven breakthroughs in precision metrology and secure information transfer (*1, 2*). However, their generation, manipulation, and measurement have historically been confined to narrowband or quasi-continuous-wave (CW) regimes. Extending squeezing into the ultrafast domain represents a critical milestone for next-generation quantum technologies, enabling access to the natural dynamical timescales of matter (*3-6*). Despite recent progress, current approaches are bottlenecked by narrowband phase-matching requirements, operation primarily in the infrared, and a lack of metrology capable of resolving the time-dependent evolution of broadband nonclassical fields.

Consequently, the generation of ultrafast bright squeezed vacuum (BSV) light has largely been restricted to up-conversion processes (*6, 7*). Unlike conventional down-conversion schemes, which struggle to sustain broadband phase-matching and temporal coherence (*8*), up-conversion enables broader phase-matching and superior pump-signal synchronization. Yet, while this pathway has successfully generated phase-squeezed ultrafast light, the realization of intensity-squeezed ultrafast pulses remains out of reach.

This limitation has profoundly impacted the emerging field of strong-field quantum optics. Recent efforts to study high-harmonic generation (HHG) from a quantum optical perspective (*9-15*) and utilize BSV pulses to control nonlinear processes (*15-27*) implicitly assume that ultrafast squeezed pulses behave similarly to traditional CW squeezed beams. This assumption is fundamentally imprecise: whereas CW squeezed fields are typically single-frequency and exhibit symmetric squeezing, ultrafast pulses are inherently broadband, encompassing a complex landscape of frequencies and intensities. Because single-frequency quantum models cannot accurately describe strong-field physics driven by broadband pulses, critical questions regarding ultrafast quantum light-matter interactions remain unresolved.

Here, we resolve these fundamental questions by establishing a bridge between quantum optics and attosecond physics. First, we uncover previously unobserved half-cycle time-dependent squeezing in ultrafast light pulses and elucidate its role in driving strong-field interactions. Second, we demonstrate attosecond control and visualization of both the amplitude- and phase-squeezed states of ultrafast light. By manipulating the nonlinear generation time window, we provide a unified physical basis for this unprecedented control. Finally, we analyze squeezed light-induced tunneling currents, demonstrating that the squeezing level of the driving light is encoded onto the induced electronic current uncertainty with attosecond precision. These capabilities establish a practical path toward ultrafast quantum metrology and real-time quantum-state engineering, forming the foundation of ultrafast quantum optics.

**Time-dependent squeezing of the ultrafast light field and its effect on strong field physics**

To generate ultrafast amplitude squeezed light, we developed a quantum light squeezer (QLS) driven by a coherent few-cycle laser pulse (~6 fs, centered at 600 nm) utilizing a degenerate four-wave mixing (FWM) process (Fig. S1A) which has emerged as a robust platform for generating ultrafast quantum light pulses (*28*); here, the broad bandwidth of the generated squeezed light demonstrates the QLS's ability to overcome the stringent phase-matching limitations typical of quantum nonlinear optics. To verify the FWM-induced squeezing beyond standard shot-noise measurements (Fig. S1B), we developed a broadband, frequency-resolved metrology to indirectly probe nonclassical noise suppression. We evaluated the intensity uncertainty ($\Delta I$)—defined as the standard deviation of the spectral intensity normalized to its mean—and compared it against the statistical baseline of the input coherent light (Fig. 1, A and B). By measuring phase fluctuations relative to a partially overlapping reference pulse, we correlated $\Delta I$ with the phase uncertainty ($\Delta\Phi$). The results (Fig. 1, C and D) confirm intensity squeezing: $\Delta I$ is suppressed relative to the coherent input, while the $\Delta\Phi$ quadrature exhibits the expected conjugate broadening. Although distinct from conventional balanced homodyne detection, this approach provides robust, experimentally accessible evidence of quadrature squeezing under well-controlled conditions.

Because broadband ultrafast squeezed light encompasses photons spanning a wide frequency range with varying intensities and spectral dispersions, individual photons are theoretically predicted to exhibit varying degrees of squeezing (*29*). Our metrology directly accesses these frequency-dependent intensity fluctuations ($\Delta I_\omega$) which we plotted in Fig. 1E. Moreover, to resolve the complementary phase dynamics, we performed 1,000 statistically time-resolved measurements (30) where we recorded the spectra as function of the delay between one of the input time respect to the other two in the FWM (as explained in SM). Using a reconstruction algorithms, we extracted the frequency-dependent phase uncertainty ($\Delta\Phi_\omega$). Figure 1E displays the retrieved $\Delta\Phi_\omega$ and $\Delta I_\omega$ alongside the pulse spectrum. Remarkably, $\Delta I_\omega$ and $\Delta\Phi_\omega$ are showing almost a conjugated opposite behavior over the broad spectrum. The frequency-resolved data reveal clear variations in $\Delta I$ across the broadband spectrum. These spectral dependencies likely arise from the nonuniform photon distribution across the ultrashort pulse spectrum combined with the wavelength-dependent nonlinear response of the dielectric medium during the FWM interaction. This variation of $\Delta I_\omega$ would implies that the $\Delta I$ of the squeezed light field is time dependent.

To investigate that further, we sampled a squeezed pulse waveform (see Fig. 1F) using the dielectric reflectivity method we established earlier (*30-32*). The $\Delta I$ of four sampled scans of squeezed light are plotted in Fig. 1G in blue and black lines, respectively. Please note, we also used the same methodology to sample classical coherent fields and show it in Fig. S2A, as a reference measurement. The results show that the intensity uncertainty of the coherent light field is random, as expected, whereas that of the squeezed light is showing a quasi-similar trend each half-cycle over the entire field (Fig. 1G). Remarkably, when zooming in on a single half-cycle of the squeezed field (see Fig. 1G inset), the squeezing level and $\Delta I$ are observed to vary within the half-cycle as a function of time. At intensity $I = 0$, $\Delta I$ is minimum (maximum phase uncertainty $\Delta\Phi$), and as the field intensity increases, $\Delta I$ increases until reaching the half-cycle intensity maximum (the phase squeezing reduced) where it shows a dip. The observed behaviors are reproduced across multiple field-sampling measurements (see Fig. S2B). Our results reveal systematic sub-cycle variations in the intensity-noise suppression of the squeezed field. These observations suggest a time-dependent redistribution of squeezing within individual optical cycles. Please note, the field sampling measurements nature might introduce certain noise which would affect the quantitively assessment. However, our observation is focus on the relative intensity uncertainty for each half-cycle within the field and compare this behavior with the coherent light field qualitatively.

The time-dependent squeezing level of the electric field significantly affects the electron dynamics in strong-field interactions such as HHG (*16*) (see Fig. 2). When the squeezing varies in time, the generated spectra and the different harmonic orders have different second-order correlation function $g^{(2)}$ as observed experimentally (*20*). To study this effect theoretically, we performed theoretical simulation HHG experiment (see Methods in SM) driven by five types of fields: (i) coherent light (reference), (ii) symmetric (time-independent) intensity-squeezed light, (iii) symmetric (time-independent) phase-squeezed light, (iv) time-dependent squeezed field 1, and (v) time-dependent squeezed field 2, with similar waveforms and cycle frequency (see Fig. S3). The corresponding HHG spectra appear in Fig. 2A. The intensity-squeezed field produce a spectrum with less cut-off the HHG spectrum coherent light field, whereas the phase-squeezed field (with large intensity fluctuations $\Delta I$) and the two time-dependent squeezed fields yield spectra with higher cutoffs and broader bandwidths. Additionally, for the time-dependent squeezing level fields 1&2, the HHG plateau are blurring (Fig. 2A) due to the fluctuation of the intensity which slightly change the electron trajectory in the continuum and its return energy. To further elucidate the underlying photon statistics nature, we calculated the $g^{(2)}$ values of the generated individual harmonics (Fig. 2B). For coherent and intensity-squeezed drivers, the odd harmonics all exhibit almost constant $g^{(2)} = 1$, compatible with coherent states of light. The time-dependent squeezed fields 1 and 2 driver yields odd harmonics with the lower orders (5, 7, 9) showing similar $g^{(2)}$ values that then the $g^{(2)}$ values vary up to $g^{(2)} = 2$ at higher harmonics, indicating super-Poissonian photon statistics. The phase-squeezed, which has the highest intensity fluctuations over the entire field half-cycles, shows significant variation of $g^{(2)}$ values over the entire odd harmonic's orders.

The time-dependent squeezing of the driven field in HHG process can be summarized as illustrated in Fig. 2C: time-dependent intensity uncertainty imprints itself onto the HHG process and is encoded in both the harmonic spectrum and photon statistics of the harmonics. The squeezed field modifies the instantaneous bending of the atomic potential, which determines the tunnel-ionization rate. Because tunneling is exponentially sensitive to the electric field, even small intra-cycle variations in intensity lead to large fluctuations in the release time and number of ionized electrons, producing strong shot-to-shot and sub-cycle variations in the electron wavepacket. If the intensity fluctuations differ between neighboring half-cycles, the potential bending becomes asymmetric with higher uncertainty of tunneling and recombination paths generating broader spectral structures, and potentially unstable attosecond emission. Such effects are expected for phase-squeezed or BSV light, which exhibit large intensity noise, whereas intensity-squeezed light—with reduced intensity fluctuations—should generate more stable XUV radiation. Remarkably, although fields 1 and 2 differ only in their time-evolving squeezing level over a half cycle, the resulting HHG spectra and the quantum statistical properties of the harmonics, $g^{(2)}$, are distinctly different. These results show that quantum uncertainty in an ultrafast squeezed field is redistributed within the optical cycle, implying that strong-field processes driven by such light cannot be described by stationary squeezing models.

**Attosecond control and visualization of quantum states of squeezed ultrafast light pulse**

In our degenerate FWM nonlinear process, three identical beams of ultrafast laser pulses—originating from a single coherent input pulse—interact to generate a nonlinear light signal. The temporal window of nonlinear signal generation (T) determines the squeezing characteristics of the generated light (see illustration in Fig. S4A). If the nonlinear generation time window is short (T is minimal), the nonlinear signal photons originate from the region of strongest temporal overlap among the three pulses. In this case, intensity fluctuations are minimized, and the generated nonlinear signal light is squeezed in the intensity quadrature, with maximum phase uncertainty in accordance with the Heisenberg principle. Conversely, if the nonlinear signal is generated over a longer time window (T is maximal), the intensity exhibits larger fluctuations (ΔI is maximal), and the light is squeezed in the phase quadrature (ΔΦ). Thus, the interaction time window T depends on the phase-matching condition, the incident angle (θ), and the relative arrival time (τ) of the three input pulses. Thus, we can control the squeezing state of the pulse with attosecond resolution by controlling $\tau$.

To demonstrate that unprecedent control we measured $\Delta I$ is at different delay time from 0 to -4 fs with 500 attosecond time steps (the estimated time jittering on our experiment setup is less than 100 as). The measurements at delays of 0, 1, 2, 3, and 4 fs are shown in Fig. S4 (B-F). These results demonstrate the change of the $\Delta I$ by changing $T$. For visualizing the attosecond control of the $\Delta I$ fluctuations, we fitted our $\Delta I$ measurements to a theoretical value $\Delta I_{th}$ based on a displaced squeezed state (*3*) of the form $|\phi\rangle = \hat{D}(\alpha)\hat{S}(r)|0\rangle$, where $\hat{D}(\alpha) = exp[\alpha\hat{\alpha}^{\dagger} - \alpha^{*}\hat{\alpha}]$ and $\hat{S}(r) = exp[\frac{r}{2}(\hat{\alpha}^{2} - \hat{\alpha}^{\dagger 2})]$ denote the displacement and squeezing operators, respectively. Here, $\alpha$ and $(\hat{\alpha})$ are the photonic creation (annihilation) operators. From the measured data, we extract the variance $\Delta I^{2}$, which we compare with the theoretically predicted variance $\Delta I_{th}^{2}(\alpha, r)$ that depends on both the coherent state intensity $\alpha$ and the squeezing parameter $r$. The optimal parameters are obtained by minimizing the cost function (see Methods)

$$C(\alpha, r) = \left|\frac{\Delta I_{th}^{2}}{\langle I\rangle^{2}} - \Delta\bar{I}\right|, \quad (1)$$

such that $$(\alpha^{*}, r^{*}) = arg[min_{\alpha,r}C(\alpha, r)].$$

Note that, consistent with the experimental definition of the uncertainties, the theoretical variance is normalized by the mean intensity $\langle I\rangle$. Using this procedure, we obtain for the optimal parameters $C(\alpha^*, r^*) \propto 10^{-15}$, which allows us to infer an effective single-mode state consistent with the measured intensity-variance statistics within the assumed displaced-squeezed-state model as (*33*)

$$W(x,p) = \frac{1}{\pi}\int_{-\infty}^{\infty} dy \langle x - y|\phi\rangle\langle\phi|x + y\rangle e^{i2py}. \qquad (2)$$

It is worth noting that this theoretical fit is restricted to a single mode, which provides a quantitatively good approximation to the properties of the generated light (*3*). This assumption is based on our pulse temporal characterization measurement shown in Fig. S1D which reveals that the pulse has a linear temporal phase indicating that our squeezing pulse is dominated by a single mode. Including additional, uncorrelated modes would correspond to replacing the variance in Eq. (1) by$\Delta I_{th}^2 = \sum_i^N \Delta I_{th,i}^2$ in Eq. (1), where $N$ denotes the total number of modes.

Accordingly, snapshots of the inferred effective Wigner representation associated with the fitted model parameters at different arrival times τ are shown in Fig. 3A. In addition, the estimated squeezing level in dB is plotted as a function of τ in Fig. 3B. By stacking all the snapshots in Fig. 3A, we generate an attosecond Wigner function movie that illustrates the real-time control of the squeezed light state (see Supplementary movie 1). The inferred effective Wigner representations are stable under the fitting procedure described, as confirmed by the analysis in the Methods section and the fitted parameters in Fig. S5. Moreover, the squeezing level of the ultrafast light pulse starts at maximum value at $\tau = 0$ fs and decreases observably for $\tau > 1$ fs as the generation time window T is increased, reaching minimum value at $\tau = 4$ fs. By tuning the relative arrival time and phase-matching geometry of the interacting pulses, we achieve attosecond-scale control over the nonlinear generation conditions that determine the effective squeezing strength and quadrature of the emitted light.

To quantify how the degree of squeezing evolves with the temporal overlap of the interacting pulses, we developed a phenomenological model that captures the dependence of the effective squeezing strength on the nonlinear generation time window T. This window represents the temporal duration over which the three optical fields overlap coherently within the nonlinear medium. As the relative delay τ between pulses changes, the nonlinear coupling strength and hence the measurable squeezing vary accordingly (see Methods in SM). Using our model, we fitted the change of squeezing level as a function of the time delay and plotted in the black line shown in Fig. 3B. The results of the fitting (as explained in SM) indicating that the sub-femtosecond modification of the arrival time would have an observable change and control of the FWM nonlinear process and the squeezing level of the generated light.

Next, we switch between intensity and phase squeezing of the generated light, we change the phase matching condition in the FWM process by changing the incident angles (θ) of the three beams on the nonlinear medium. Then we performed the statistic measurements of the $\Delta I$ and the phase uncertainty ($\Delta\Phi$) (similar to the measurement done in Fig. 1A-D). The measurements of $\Delta I$ and $\Delta\Phi$ are shown in Fig. 4A-D, respectively. The statistical data reveal a clear trade-off between intensity and phase uncertainties and that the phase is squeezed, and intensity is anti-squeezed, confirming the tunability and utmost control of the squeezed light quantum state. In time domain, when the phase-matching angle θ is set to optimize phase squeezing, we study the underlying dynamics by measuring the $\Delta I$ as a function of the arrival time delay. Please note, the phase uncertainty ($\Delta\Phi$) cannot be traced at different delay steps since this would introduce a phase modulation not related to the $\Delta\Phi$ jittering. Hence, the change is indirectly investigated by measuring $\Delta I$. A weak intensity squeezing is observed at perfect temporal overlap of the three beams ($\tau = 0$ fs). This intensity squeezing enhances as the arrival time $\tau$ increases since this delay compensate for the difference in the arrival time between the three pulses due to the change of the phase matching angle and the tilt of the nonlinear medium. The obtained snapshots of Wigner function at different time $\tau$, from fitting the measured $\Delta I$ (following the same approach as in Fig. 3) are shown in Fig. 4E. The movie is provided in supplementary movie S2. The intensity squeezing level starts at the lowest level at $\tau= 0$ fs, then increases at higher $\tau$ until reaches maximum squeezing at $\tau = 4$fs. These results show the femtosecond control of the squeezing states for the phase squeezed ultrashort laser pulses and supported with our fitting curve shown in black line in Fig. 4F (see Methods and SM). These results demonstrate attosecond control of the quantum state of light, advancing our understanding of ultrafast squeezed-light physics.

**Encoding the quantum state of light on photoinduced current**

A central goal of ultrafast quantum optics is to elucidate the interaction between quantum light and matter. Building on our recent demonstration of petahertz electronic switching using classical optical fields (*34*), we utilize ultrafast intensity-squeezed light pulses to drive electron tunneling in a similar graphene-silicon-graphene heterostructure (Fig. 5A). We simultaneously measure the optical intensity uncertainty ($\Delta I$) of the squeezed drive light and the resulting tunneling current uncertainty ($\Delta J$) across various squeezing levels, which we control by varying the delay in 500-attosecond steps (Fig. S6). The ensemble statistics (Fig. 5B) reveal a clear correlation: the uncertainty of the tunneling current ($\Delta J$) directly scales with the optical intensity statistics ($\Delta I$) of the nonclassical drive field.To definitively rule out power-dependent artifacts, we measured the current statistics while varying the pump power at a fixed squeezing level ($T = 0$ fs). As shown in Fig. S7, $\Delta J$ is dominated by electronic and thermal noise at low powers and scales up as power increases. This behavior starkly contrasts with our primary observation in Fig. 5B, where $\Delta J$ is minimized at $T = 0$ fs precisely when the squeezed light power is at its maximum. This confirms that the observed variations in current noise genuinely encode the quantum state of the light, rather than simply reflecting fluctuations in the induced optical power. Furthermore, these results establish our petahertz quantum current phototransistor (*34*) as an ultrafast quantum sensor capable of attosecond temporal resolution.

In summary, we have experimentally demonstrated an ultrafast quantum light field squeezer capable of directly generating and actively controlling squeezed states across an ultrabroad spectral range. Our frequency-domain metrology reveals that the intensity uncertainty of squeezed light varies temporally within the optical field, fundamentally influencing strong-field interactions such as high-harmonic generation (HHG). The observation of intra-cycle, time-dependent squeezing provides a new framework for understanding the real-time evolution of nonclassical light on its natural timescale. Furthermore, we provide an experimentally accessible route to the sub-femtosecond tuning of effective squeezing characteristics, offering critical new insights into HHG, entanglement dynamics, and ultrafast decoherence. Crucially, by demonstrating that nonclassical ultrafast light can dictate the noise properties of a tunneling current, we establish a foundational hybrid quantum-electronic interface where macroscopic electronic signals directly inherit the quantum features of a driving optical field. Looking forward, this methodology unlocks unprecedented opportunities for ultrafast quantum metrology, highly sensitive time-resolved spectroscopy, and next-generation quantum communication and computation. By effectively bridging ultrafast optics and quantum technologies, these results lay the groundwork for a transformative new era of ultrafast quantum optoelectronics.

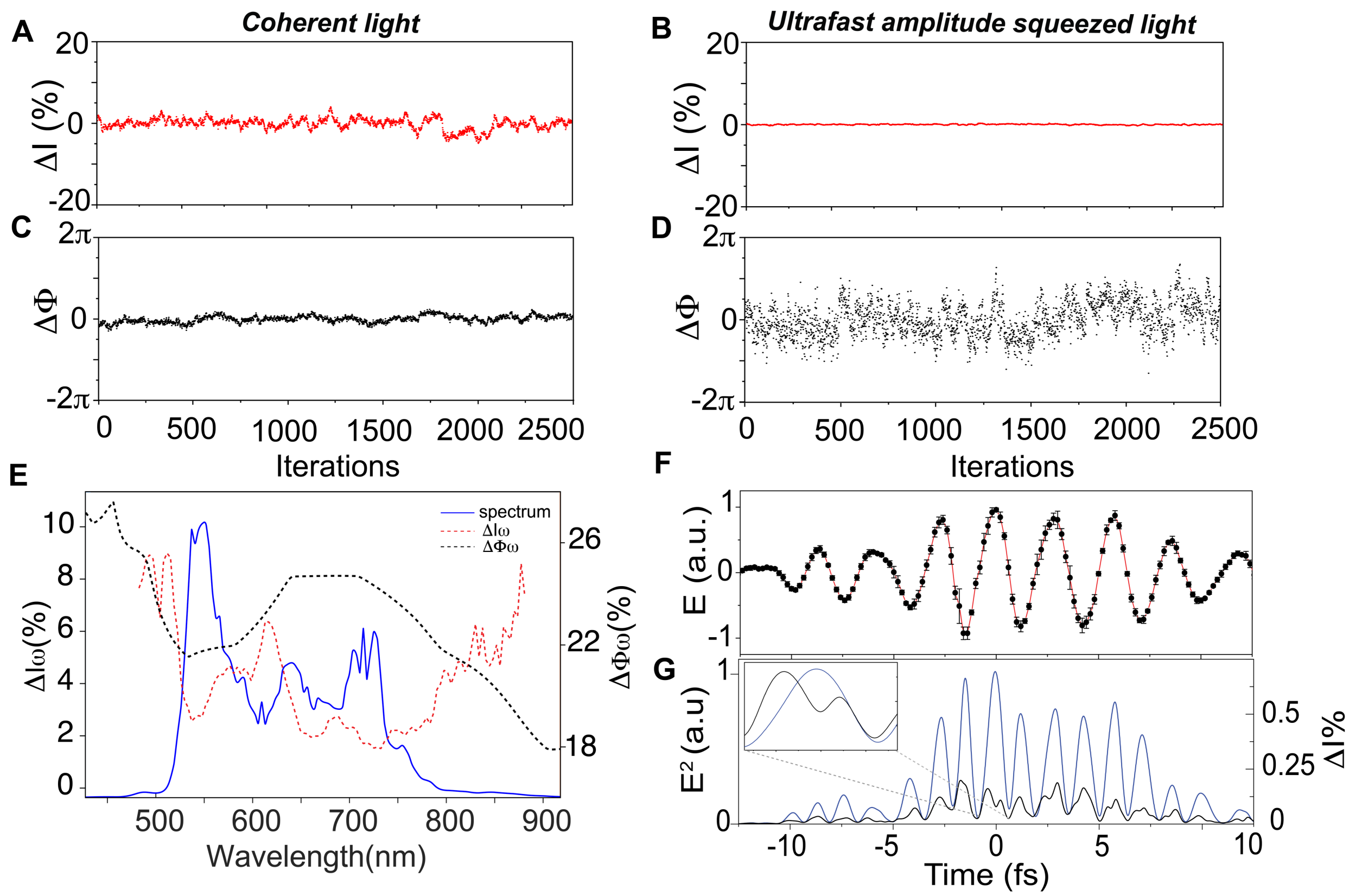

**Figure 1 | Frequency and time-dependent squeezed ultrafast light field. (A-D)** The measured $\Delta I$ and $\Delta\Phi$ for the classic (**A** &**C**) and squeezed light (**B** &**D**). **(E)** Measured frequency-dependent intensity (dashed red line) and phase (dashed black line) uncertainty across the broadband spectrum (blue line) of the ultrafast squeezed pulse. **(F)** Average of four sampled squeezed-light fields, with error bars indicating the standard deviation of the field intensity. **(G)** Field intensity and the corresponding relative $\Delta I$ of the squeezed field. The inset zooms into a single half-cycle, revealing both the asymmetry and the time-dependent evolution of $\Delta I$ within that half-cycle.

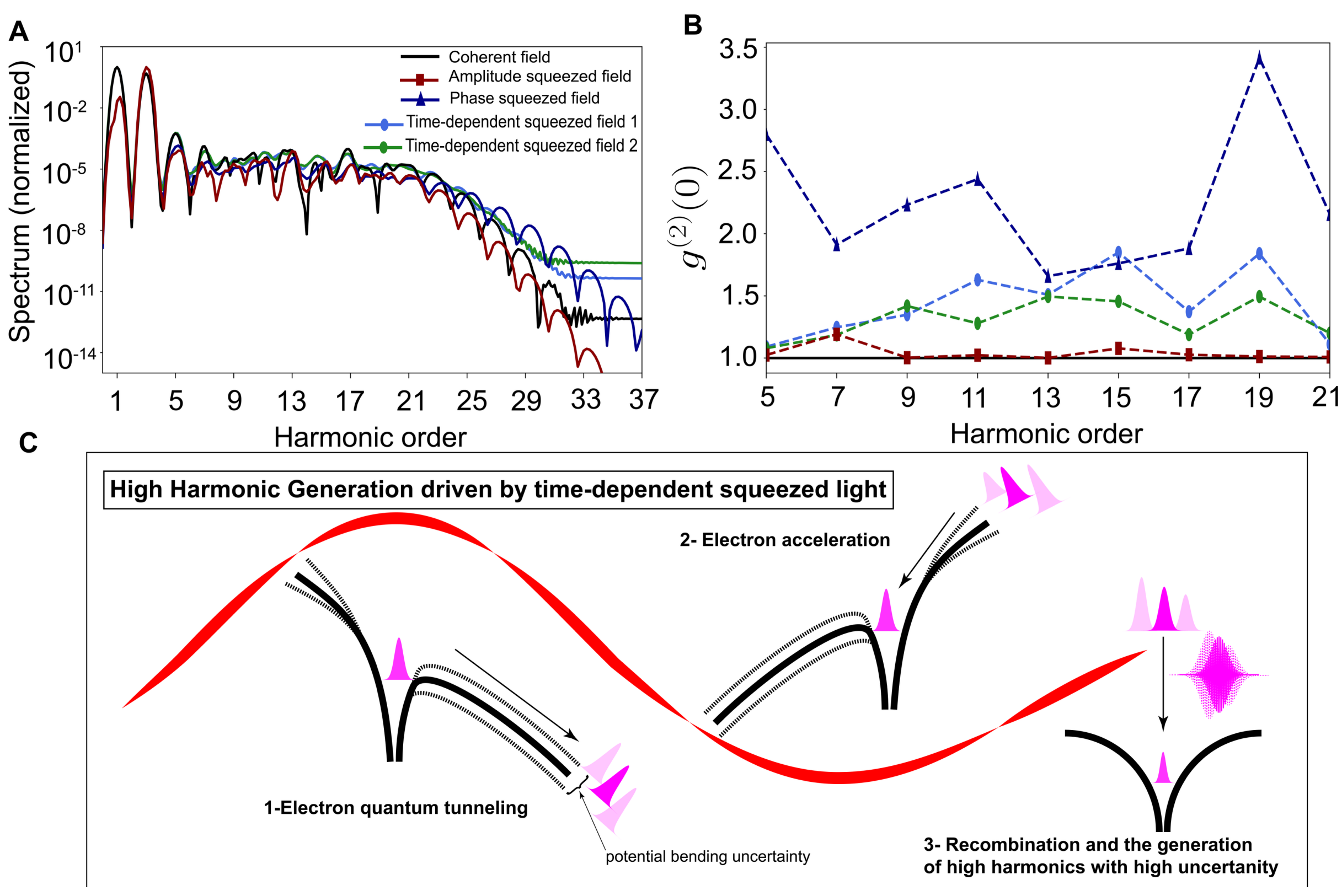

**Figure 2 | Strong field high harmonic generation driven by time-dependent squeezed light field. (A-B)** The calculated HHG spectra and $g^{(2)}$ driven by coherent, intensity squeezed, phase squeezed, time-dependent squeezing field 1 and 2. **(C)** Schematic illustrating how sub-half-cycle variations in the squeezing level of a driver pulse influence strong-field interactions (e.g. HHG), including both the squeezing and the trajectories of the resulting high harmonics.

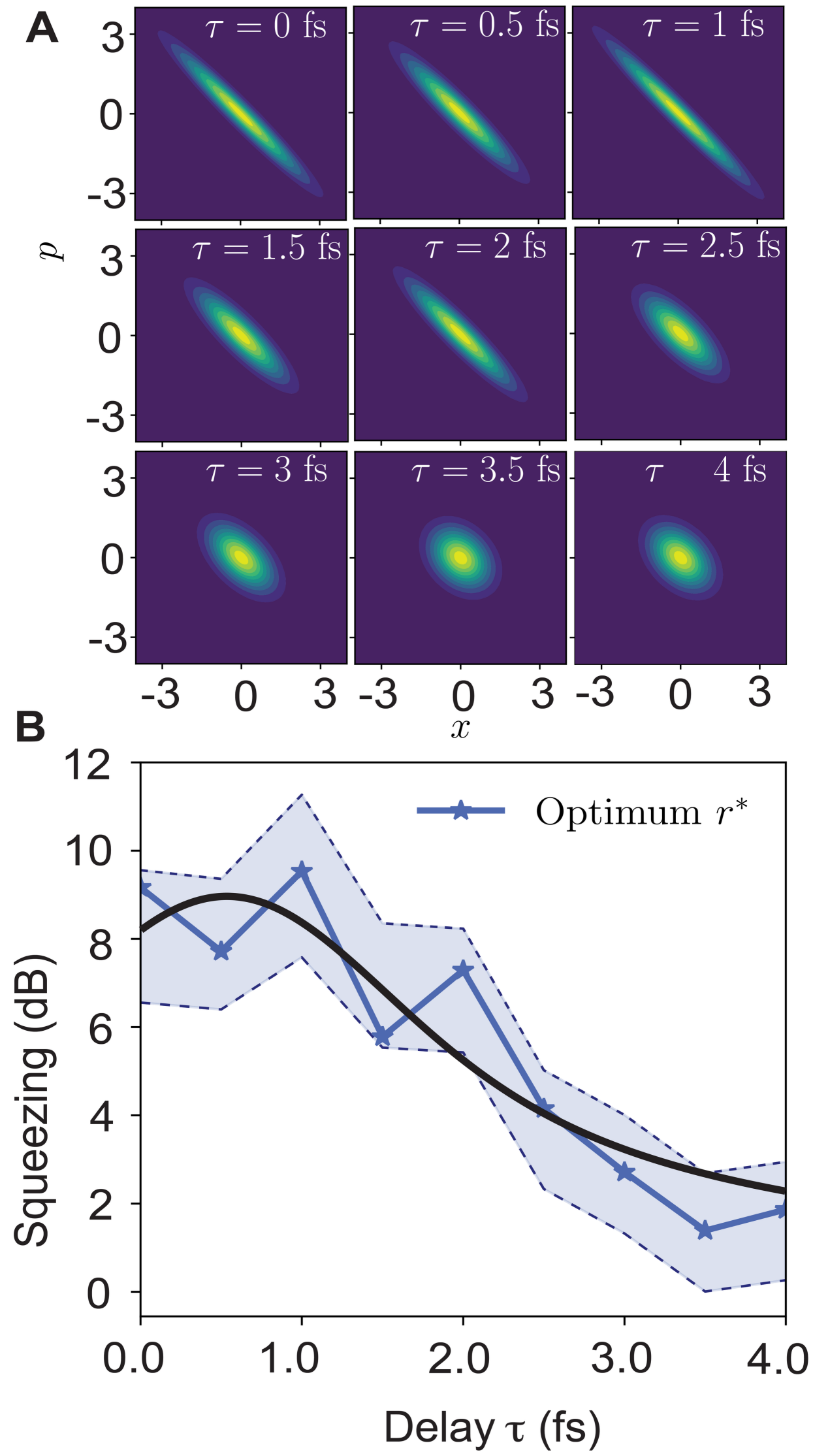

**Figure 3 | Attosecond control and visualization of the Wigner-function dynamics of an intensity-squeezed ultrafast light pulse. (A)** Snapshots of the obtained Wigner functions at different delays, recorded from 0 to 4 fs in 500-attosecond steps, showing the continuous evolution of the squeezed quantum state. In these plots, x and p represent the optical quadrature, and for representational purposes, the distributions have been rotated by 45º relative to the p-axis. The full attosecond Wigner-function movie is provided as Supplementary Movie S1.(**B**) Extracted squeezing level as a function of delay τ, revealing attosecond-precision control over the intensity quadrature. The blackline presents the fitting of the results using the phonemical model explained in SM.

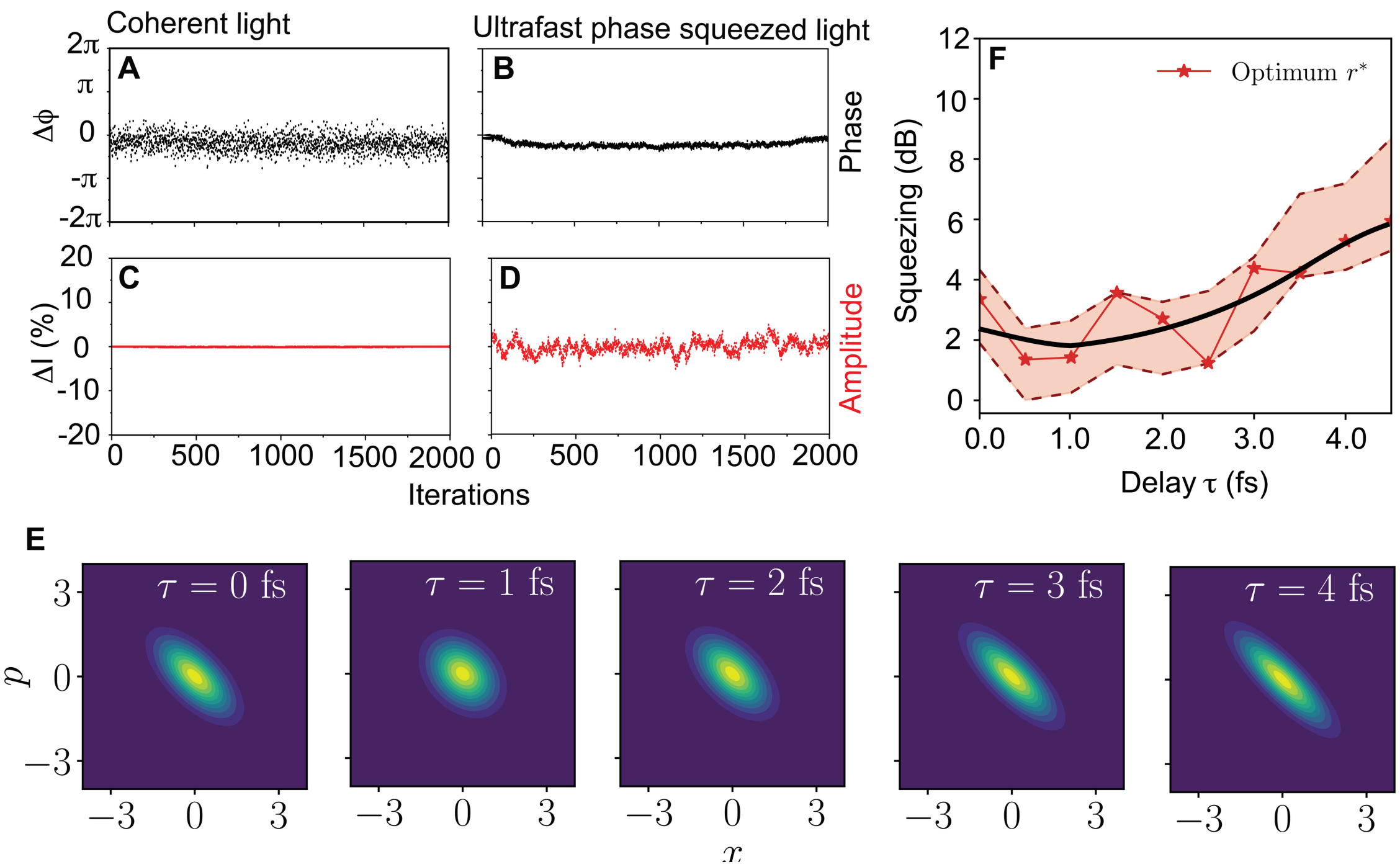

**Figure 4 | Attosecond quantum uncertainty dynamics of phase-squeezing dynamics of ultrafast pulses.** (**A-D**) Measured phase variance (ΔΦ) and intensity variance (ΔI) for coherent light (**A & C**) and phase-squeezed light (**B, D**). The reduced phase noise and corresponding increase in intensity noise in panels b and d confirm active switching to phase-quadrature squeezing by tuning the phase-matching angle in the QLS. **(F)** Representative snapshots of the obtained Wigner functions at different temporal delays τ, demonstrating femtosecond-scale control over the phase-squeezed quantum state. $x$ and $p$ denote the optical quadratures, and for illustrative purposes, the distributions have been rotated by 45° relative to the $p$-axis. The corresponding squeezing level retrieved is shown in **E.** The black line is the fitting of the squeezing level variation as a function of time.

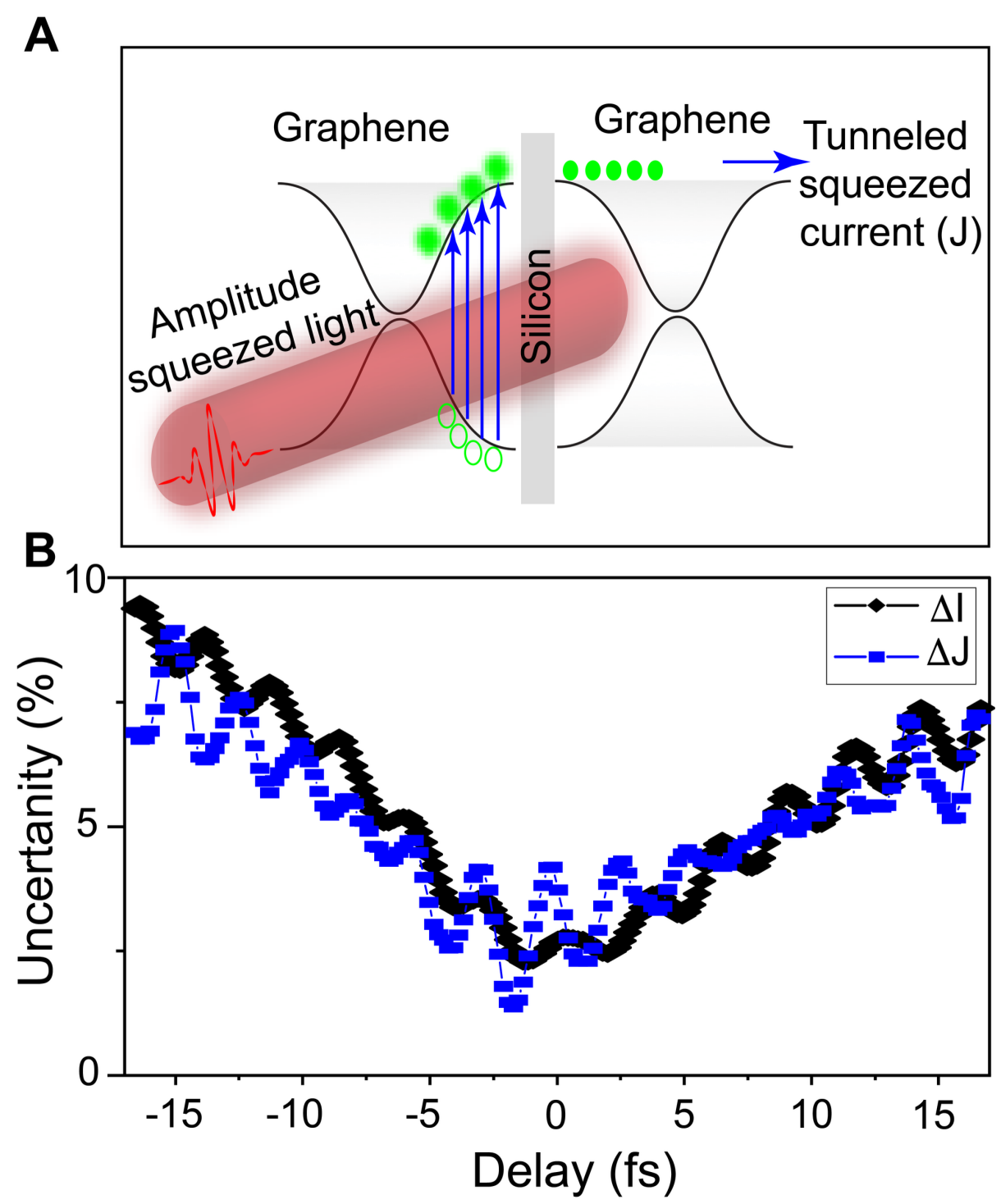

**Figure 5 | Attosecond quantum light state encoding on photoinduced tunneling current (A)** Illustration of the quantum coupling between the driver squeezed light and the generated quantum tunneling current in Graphene-Silicon-Graphene phototransistor. **(B)** The static measurements (average of 100 points) of the squeezed light intensity ($\Delta I$) and the induced-tunneling current ($\Delta J$) uncertainties measured simultaneously.